\documentclass[%
reprint,
superscriptaddress,
amsmath,
amssymb,
pra,
floatfix,
]{revtex4-1}

\usepackage{lipsum}
\usepackage{lmodern}
\usepackage{hyperref}
\usepackage[]{algorithm2e}
\usepackage{color}
\usepackage{algpseudocode}
\usepackage{siunitx}

\DeclareSIUnit{\belmilliwatt}{Bm}
\DeclareSIUnit{\dBm}{\deci\belmilliwatt}

\usepackage{import}
\usepackage{amsmath}

\usepackage{graphicx}
\usepackage{dcolumn}
\usepackage{bm}

\newcommand{\Var}{\operatorname{Var}} 

\begin{document}
\graphicspath{{./figures/}}

\title{Automated long-range compensation of an rf quantum dot sensor}

\author{Joseph Hickie}
\thanks{These authors contributed equally.}
\affiliation{Department of Materials, University of Oxford, Oxford OX1 3PH, United Kingdom}

\author{Barnaby van Straaten}
\thanks{These authors contributed equally.}
\affiliation{Department of Materials, University of Oxford, Oxford OX1 3PH, United Kingdom}

\author{Federico Fedele}
\thanks{These authors contributed equally.}
\affiliation{Department of Materials, University of Oxford, Oxford OX1 3PH, United Kingdom}

\author{Daniel Jirovec}
\affiliation{Institute of Science and Technology Austria, Am Campus 1, 3400 Klosterneuburg, Austria}

\author{Andrea Ballabio}
\affiliation{L-NESS, Physics Department, Politecnico di Milano, via Anzani 42, 22100, Como, Italy}

\author{Daniel Chrastina}
\affiliation{L-NESS, Physics Department, Politecnico di Milano, via Anzani 42, 22100, Como, Italy}

\author{Giovanni Isella}
\affiliation{L-NESS, Physics Department, Politecnico di Milano, via Anzani 42, 22100, Como, Italy}

\author{Georgios Katsaros}
\affiliation{Institute of Science and Technology Austria, Am Campus 1, 3400 Klosterneuburg, Austria}

\author{Natalia Ares}
\affiliation{Department of Engineering Science, University of Oxford, Oxford OX1 3PJ, United Kingdom}

\email{natalia.ares@eng.ox.ac.uk}
\date{\today}

\date{\today}

\begin{abstract}
Charge sensing is a sensitive technique for probing quantum devices, of particular importance for spin qubit readout. To achieve good readout sensitivities, the proximity of the charge sensor to the device to be measured is a necessity. However, this proximity also means that the operation of the device affects, in turn, the sensor tuning and ultimately the readout sensitivity. 
We present an approach for compensating for this cross-talk effect allowing for the gate voltages of the measured device to be swept in a \SI{1}{\volt} $\times$ \SI{1}{\volt} window while maintaining a sensor configuration chosen by a Bayesian optimiser.
Our algorithm is a key contribution to the suite of fully automated solutions required for the operation of large quantum device architectures.
\end{abstract}

\maketitle

\section{Introduction}

Radio-frequency (rf) charge sensing using a proximal quantum dot has become a ubiquitous technique for fast and sensitive measurements of quantum devices~\cite{Schoelkopf1998, Barthel2010FastDot, Ares2016, Muller2010a, Noiri2020a, VigneauProbing}. This technique is used for single-shot readout of spin qubits owing to its high sensitivity~\cite{West2019, Hogg2023}. As devices scale to multi-qubit architectures, the number of readout sensors per chip is bound to increase ~\cite{Philips2022, Fedele2021}, making their manual tuning and optimisation an increasingly time-consuming task. 

The highest sensitivities provided by rf charge sensing rely on the high transconductance of the sensor dot when tuned to the flank of a Coulomb peak. But the cross-talk between the device and the sensor dot results in this regime being quickly lost when sweeping the device gate voltages. Techniques based on digital feedback have been developed to compensate for this cross-talk effect~\cite{Yang2011a, Nakajima2021a}. These approaches rely on maintaining a constant current flowing through the sensor dot, but do not necessarily maintain a high transconductance, using available Coulomb peak flanks irrespective of their conductance gradient. These methods are thus not able to keep the readout contrast, i.e. the sensor's  sensitivity, constant as the device gate voltages are swept through ranges over $100 \times 100$ \si{\milli\volt}.

In this paper, we introduce an automated approach to find and maintain the high sensitivity of an rf sensor dot while sweeping across a large range of a device's gate voltage space. The algorithm is demonstrated in a hole-based Ge/SiGe heterostructure quantum dot array but is agnostic to the device material and geometry. Our algorithm employs Bayesian optimisation to optimally tune the rf dot to a regime where it is highly sensitive. Once the optimal readout contrast is found, virtual gates~\cite{Hensgens2017, Botzem2018, Volk2019, Mills2019} are used to maintain the readout sensitivity in fast $100 \times 100$ \si{\milli\volt} measurement windows. This ac-based compensation is combined with a dc-based compensation algorithm to allow for the rf sensing of a stability diagram over \SI{1}{\volt} $\times$ \SI{1}{\volt}, in which we can distinguish more than $25 \times 25$ charge transitions, using a single sensing peak. Our compensation algorithm can be repeatedly recalibrated to arbitrarily extend this range, thus overcoming the limitation of the linear approximation required by the constant interaction model.

\begin{figure}
\centering\includegraphics{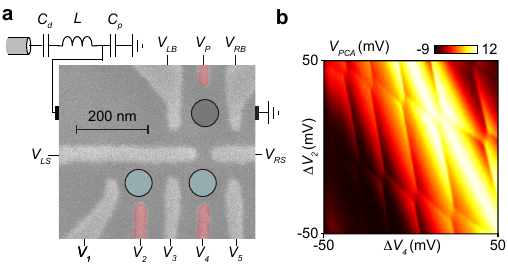}
  
  \label{fig:setup}
  \caption{\textbf{a}) A scanning electron microscopy image of a device nominally identical to the one used in this experiment. The gate electrodes connected to microwave lines are colour-coded in pink. A bias tee combines a dc offset with the ac signal for the gates in red false colour. \textbf{b}) Demodulated signal from the rf-QD obtained by PCA when the device is in a double dot regime. For a given point in gate voltage space, we perform this 2D measurement by ramping gates $V_2$ and $V_4$ in a raster pattern with an arbitrary waveform generator.}
\end{figure}

\section{Device and measurement setup}
\label{sec:setup}

These experiments were performed in an electrostatically-defined Ge/SiGe quantum dot (QD) fabricated in a similar manner to that described in \cite{Jirovec2021}. Gates $V_1 - V_5$ are used to define quantum dots in the main device, while gates $V_\mathrm{LB}, V_\mathrm{P}, $ and $V_\mathrm{RB}$ define the charge sensor. Gates $V_\textrm{LS}$ and $V_\textrm{RS}$ separate the sensing dot from the device dots, and were both set to \SI{1.8}{\volt} for this experiment.  For readout, we connected the ohmic of the charge sensor to an L-matching network with a \SI{92}{\pico\farad} decoupling capacitor $C_\mathrm{d}$, \SI{2.7}{\micro\henry} inductor $L$, and a parasitic capacitance $C_\mathrm{p}$ (Fig.~\ref{fig:setup}) \cite{Ares2016}. We used a Zurich Instruments UHFLI to drive the circuit at a power of \SI{-90}{\dBm} and demodulate the reflected signal.  

The presented algorithm relies on fast 2D measurements taken by ramping the voltage on gates $V_2$ and $V_4$ in a raster pattern using an arbitrary waveform generator (AWG) \cite{Stehlik2015, Schupp2020, vanStraaten2022a}. In these 2D measurement windows, the output waveforms are compensated to minimise the effect of distortion introduced by the filters as in \cite{vanStraaten2022a}. We measured the demodulated in-phase (\textit{I}) and out-of-phase (\textit{Q}) components of the reflectometry signal from the rf-QD sensor for each pixel with an integration time of \SI{1}{\micro\second} (Appendix~\ref{appendix:measurements}). To optimally capture both the \textit{I} and \textit{Q} components of the rf-QD signal, we use principal component analysis (PCA). The signal of the rf-QD demodulated in this way is labelled $V_{\mathrm{PCA}}$ and corresponds to the first principal component of the measured data. For more details on the use of PCA, see Appendix~\ref{appendix:pca}.

\section{Algorithm}

Our algorithm consists of two stages. The first stage optimally tunes the charge sensor barriers ($V_\mathrm{LB}, V_\mathrm{RB}$) to achieve maximum sensitivity (Section~\ref{sec:bayesopt}). Optimal barriers are found when the rf-QD's impedance falls within a specific range such that the readout matching circuit has an impedance close to the $Z_0 = $ \SI{50}{\ohm} line impedance. We use a direct approach by defining a score function for the visibility of the charge sensor and then optimise it using Bayesian optimisation.

The second stage is designed to maintain the tuning of the charge sensor while the device gate electrodes ($V_{1-5}$) are swept (Section~\ref{sec:automated_decoupling}). We compensate for this crosstalk by constructing virtual gates. These virtual gates include the charge sensor's plunger gate ($V_\mathrm{P}$) to compensate for the capacitive coupling between the device's gate electrodes and the rf-QD sensor.

\subsection{Optimally tuning charge sensor tunnel barriers using Bayesian optimisation}
\label{sec:bayesopt}

\subsubsection{Variance-based score function for sensor visibility}
\label{sec:score_function}

To optimise the rf-QD sensor's sensitivity, we define a score function. With optimally-tuned barriers, the difference between the reflected signal on and off a Coulomb peak is maximised, maximising the height of the peaks and thus the readout sensitivity. We have found that maximising the variance of a charge sensing measurement is a fast and noise-resistant method for optimising the visibility of the charge sensor (Appendix~\ref{appendix:var}). 

To evaluate the score function, we initially take a fast 2D measurement as described in Section~\ref{sec:setup} and determine the variance of the data after projection onto the first principal component, $\Var(V_{\mathrm{PCA}})$. 
The variance is a natural choice for a score function, and can be used to optimise not just the position of the Coulomb peak but also many other experimental parameters such as rf frequency and power (Appendix~\ref{appendix:optimising_parameters}). The variance of $V_{\mathrm{PCA}}$ is conceptually the same as considering the variance of the demodulated \textit{I} component in a measurement where the phase is perfectly tuned (Appendix~\ref{appendix:pca}). If we only had access to the \textit{I} component of the demodulated data, the score function could be evaluated by taking the variance in the \textit{I} component. In this case, the phase would need to be optimised as a parameter in the subsequent Bayesian optimisation. 

\begin{figure}
\centering
  \includegraphics{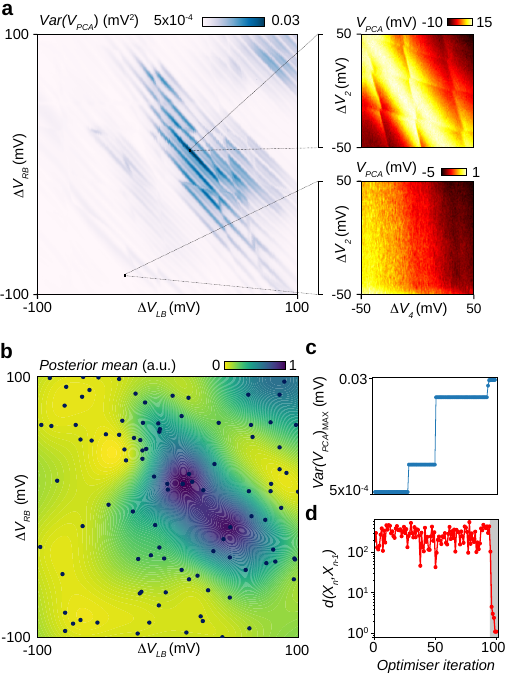}

  \caption{\textbf{a}) An example of the underlying score function over which we run the barrier optimisation routine, representing the variance of $\mathrm{V_{PCA}}$ for a \SI{200}{\milli\volt} $\times$ \SI{200}{\milli\volt} window in the gate voltage space given by the $V_{\mathrm{LB}}$ and $V_{\mathrm{RB}}$. Each pixel in this scan represents the variance of $\mathrm{V_{PCA}}$ in a fast 2D measurement. The insets shows examples of these measurements corresponding to two pixels with high and low variance. 
  In regions where the rf-QD is sensitive, the variance is high (upper inset). In regions where the rf-QD is not sensitive, the variance is low as the measurement is primarily noise (lower inset). \textbf{b}) 
 The posterior of a Gaussian process over a set of values of $\Var(V_\textrm{PCA})$ from \textbf{a}, marked by black points and chosen by the optimiser. The optimiser aims to find the maximum value of $\Var(V_\textrm{PCA})$ in the gate voltage window considered. \textbf{c}) The maximum value of $\Var(V_\textrm{PCA})$ that was found by the optimiser as a function of the number of iterations. As the optimisation progresses, the optimal value of $\Var(V_\textrm{PCA})$ increases. \textbf{d}) Distance between sequential gate voltage coordinates chosen by the optimiser. This distance decreases significantly for the last few iterations of the optimiser (grey area).
 }\label{fig:barrier_bayes}
\end{figure}

\subsubsection{Bayesian optimisation of rf-QD sensor barrier gates}
\label{sec:bayesian_optimisation}

We aim to optimise our readout sensitivity by controlling the transparency of the rf-QD's tunnel barriers through gates $V_\mathrm{LB}$ and $V_\mathrm{RB}$. Our Bayesian optimiser proposes barrier gate voltages at which to evaluate the score function. In  Fig.~\ref{fig:barrier_bayes} \textbf{a}, we show an example of the gate voltage parameter space over which we optimise the barriers. Each pixel in this figure panel is the variance of V$_\mathrm{PCA}$ for a fast 2D measurement. To find the optimal barrier voltage, we could perform this measurement and take the values of V$_{LB}$ and V$_{RB}$ that result in $\max \Var(V_\mathrm{PCA})$. However, this is a very time-consuming approach, considering this measurement is made up of $500 \times 500$ fast 2D measurements. In our optimisation approach, V$_{LB}$ and V$_{RB}$ are chosen by the optimiser. Initially, these voltages are chosen to explore the domain of the score function, building the Gaussian process model (Fig.~\ref{fig:barrier_bayes} \textbf{b}). Later iterations exploit the information in the model to converge on the global maximum (Fig.~\ref{fig:barrier_bayes} \textbf{c}, \textbf{d}). A sharp Coulomb peak, optimal for sensing, will be thus found at the coordinates in gate voltage space which correspond to the global maximum of $\Var(V_\textrm{PCA})$. Optimising the barriers in this way requires approximately 100 fast 2D measurements. 

We use a Bayesian optimiser with a Mat\'ern 5/2 kernel and length scale constrained to be greater than the Coulomb peak spacing. This forces the model to only capture the large-scale variations in the score function, without the short-scale periodic variation. We found that making the optimiser blind to the score function's periodicity in this way allowed it to efficiently find the global maximum. 

Our Bayesian approach is motivated by the fact that the score function landscape displayed in  Fig.~\ref{fig:barrier_bayes} \textbf{a} defeats standard optimisers such as Nelder-Mead or BFGS. We hypothesise that this is due to the complex, periodic nature of the optimisation landscape given by Coulomb peaks in the rf-QD.

\subsection{Compensating for crosstalk}
\label{sec:automated_decoupling}

In the previous stage of the algorithm, we have identified an optimal Coulomb peak for sensing. This Coulomb peak, present in the rf-QD sensor, is affected by the device gate voltages due to crosstalk. We aim to create virtual gates for the quantum dot device that allows us to move through the devices' gate voltage space while keeping the rf-QD Coulomb peak fixed in the 2D measurement window. We achieve this by compensating each device gate with a contribution from $V_\mathrm{P}$. The resulting virtual gates $V^{'}_{1-5}$ would necessarily require $V_\mathrm{P}$ to act in the opposite direction to the device gate.

\begin{figure}
\centering
  \includegraphics{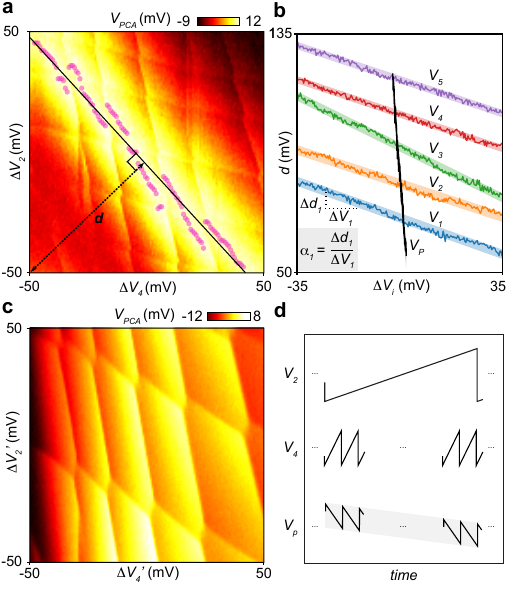}
  \caption{\textbf{a}) A 2D measurement window. Purple points identify the rf-QD Coulomb peak maximum following the method described in Appendix~\ref{appendix:coulomb_fitting}. The solid line is a regression fit through these points. The dashed line indicates \textit{d}, the perpendicular distance of the rf-QD Coulomb peak to the lower left corner of the 2D measurement window. 
   \textbf{b}) Values of the Coulomb peak offset, $d$, as a function of changes in gate voltage for each of the gates of interest. Note the curves have been vertically offset to aid the reader. The sensor's plunger gate, $V_\mathrm{P}$, has the strongest gradient, as expected. \textbf{c}) Example of a 2D measurement window after ac compensation. The uncompensated version can be found in Fig. 1 \textbf{b}. \textbf{d}) The top two waveforms are schematics of the voltages applied to device gates to perform a 2D measurement window. The bottom waveform is the ac compensation voltage applied to $V_\mathrm{P}$.}\label{fig:d_measurements}
\end{figure}

We begin by measuring the effect of each gate $V_{1-5}$ and $V_\mathrm{P}$ on the position of the rf-QD Coulomb peak in the 2D measurement window (Fig.~\ref{fig:d_measurements} \textbf{a, b}). This position ($d$) is defined to be the shortest distance from the lower left corner of the 2D measurement to a linear regression fit of the rf-QD Coulomb peak  (Appendix~\ref{appendix:coulomb_fitting}). We quantify the strength of the effect of the gate voltages on $d$ as $\alpha_i = \Delta d_i / \Delta V_i$, where $\Delta d_i$ is the change in $d$ when the device gate voltage $V_i$ is changed by a given $\Delta V$. We can choose a gate voltage range $\Delta V_i$ within which the values of $\alpha_i$ are constant, i.e.\ there is a linear dependence 
between $d$ and $V_i$. If the full width of the rf-QD Coulomb peak is not visible in the 2D measurement window, the estimation of $d$ becomes less reliable. An alternative approach to estimate the strength of the effect of the gate voltages on the position of the rf-QD Coulomb peak is to measure the finite difference gradient of the peak edges for each of the gates $V_{1-5}$ and $V_\mathrm{P}$ (Appendix~\ref{appendix:finite_differences}). 

The strength of the compensating contribution from $V_\mathrm{P}$ in each virtual gate $V_i^{'}$ is given by a coefficient $\gamma_i$, given by $\gamma_i = \alpha_i / \alpha_P$, where $\alpha_P$ is the strength of the effect of $V_\mathrm{P}$ on $d$. The values of $\gamma_i$ quantify the relative strength of each device gate compared to the strength of $V_\mathrm{P}$ on $d$. We would expect that $\gamma_i$ is thus less than 1 since the plunger gate has a greater effect on the rf-QD Coulomb peak position than any of the device gates.

The time taken to estimate $\gamma_i$ depends on several factors but is entirely deterministic. These factors include the resolution of the 2D measurement windows, the pixel integration time, the time it takes to sweep a gate voltage over $\Delta V_i$, the processing speed of the computer, the number of device gates, and the number of sensors being compensated. In this setup, with five device gates and a single sensor plunger, the time taken to generate all the values of $\gamma_i$ is approximately 10 seconds. This
time could be reduced at the expense of noise robustness. 

Our cross-talk compensation algorithm extracts the values of $\gamma_i$ to achieve compensation over long-range device gate sweeps and for the gate sweeps within a 2D measurement window. We will refer to the former (latter) as dc (ac) compensation. 

\subsubsection{Dc compensation}

As we perform a long-range device gate sweep, we navigate the device's stability diagram. The position \textit{d} of the rf-QD Coulomb peak in the measurement window changes as a result of this sweep.
To compensate for a change in voltage $\Delta V$ on device gate $i$, the compensated device gate $V_i^{'}$ must increment the voltage on the sensor plunger gate $V_\mathrm{P}$ by a value opposite in sign and scaled by the relative strength of the compensating contribution, $- \gamma_i \Delta V$. This compensated device gate has a unit vector defined by

\begin{equation}
    \hat{V}_i' \propto (\hat{V}_i - \gamma_i \hat{V}_\mathrm{P}). 
    \label{eq:single_compensated_gate}
\end{equation}

To generalise this concept, we first define an uncompensated virtual gate, which has an arbitrary direction in gate voltage space spanned by $\{V_1, V_2, V_3, V_4, V_5, V_P\}$: 

\begin{equation}
     \hat V_\textrm{arb} \propto [\epsilon_1, \epsilon_2, \epsilon_3, \epsilon_4, \epsilon_5, 0]^T, 
\label{eq:arbitrary_uncompensated_virtual}
\end{equation}

where $\epsilon_i$ represents the contribution of gate $i$ to the virtual gate. 

The corresponding compensated virtual gate is constructed by including the contributions to $V_P$: 

\begin{equation}
     \hat V_\textrm{arb}' \propto \left[\epsilon_1, \epsilon_2, \epsilon_3, \epsilon_4, \epsilon_5, - \sum_{i=1}^5 \gamma_i \epsilon_i \right]^T.
\label{eq:compensated_arbitrary_virtual}
\end{equation}

\subsubsection{Ac compensation}
\label{sec:fast_compensation}
Within a measurement window, we can observe the rf-QD Coulomb peak due to the presence of cross-talk.

To compensate for this crosstalk, we create a waveform to compensate the fast sweeps that define the 2D measurement window. This third waveform is a linear combination of the two waveforms being used to generate the plunger fast sweeps (Fig.~\ref{fig:d_measurements} \textbf{c, d}): 

\begin{equation}
\label{eq:VPAC}
V_\mathrm{P}(t)= -\begin{bmatrix}
\gamma_2 \\ 
\gamma_4 \\
\end{bmatrix} \cdot \begin{bmatrix}
V_2(t) \\
V_4(t)
\end{bmatrix}
\end{equation}

where $V_\mathrm{P}(t), V_2(t), V_4(t)$ are the ac components of the voltages applied to each of the three gates. $V_\mathrm{P}$ compensates for the cross-capacitive coupling between the device gates and the rf-QD. In this way, we remove the gradients corresponding to the rf-QD Coulomb peak from the 2D measurement window background (Fig.~\ref{fig:setup} \textbf{d}).
This ac compensation approach is readily applicable to arbitrary pulses involving any number of device gates. Equation \ref{eq:VPAC} is thus generalised by including additional time-dependent components scaled by the relevant $\gamma$ value.

\begin{figure}
\centering
  \includegraphics{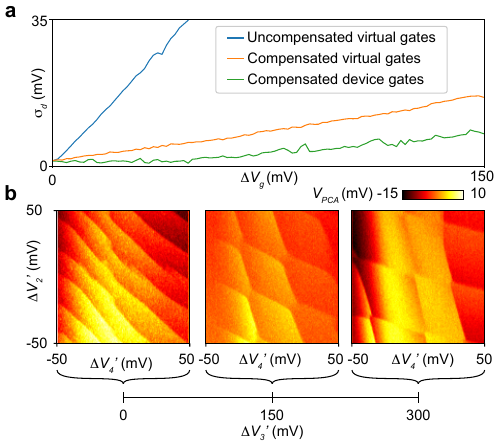}
  \caption{\textbf{a}) Standard deviation of a Coulomb peak's position quantified by $d$, $\sigma_d$, as a function of the length of the gate voltage sweep, $\Delta V_g$, for uncompensated virtual gates (as per Eq.~\ref{eq:arbitrary_uncompensated_virtual}), and for a different combination of gates in the compensated case (as per Eqs.~\ref{eq:single_compensated_gate}, \ref{eq:compensated_arbitrary_virtual}). \textbf{b}) Three 2D measurement windows for different gate voltage configurations. These configurations are accessed by changing the value of the compensated middle barrier gate voltage, $V_3'$, by $\Delta V_3'$. Our compensation algorithm allows us to choose a Coulomb peak of our sensor dot and use it to explore the device gate voltage space going from a single dot to a double dot regime. Ac compensation is applied for each of these 2D measurement windows.}\label{fig:performance}
\end{figure}

\section{Results}

As discussed in Section~\ref{sec:automated_decoupling}, a perfectly-compensated gate $V_\mathrm{g}'$ would have an associated $\alpha_{\mathrm{g}} = 0$, meaning the Coulomb peak offset $d$ does not change as the compensated gate voltage $V^{'}$ is modified. We therefore measure the performance of our dc compensation algorithm by calculating the standard deviation of $d$, $\sigma_d$, as a function of gate voltage sweeps for three cases. In the first case, we generate uncompensated virtual gate sweep directions as per Equation~\ref{eq:arbitrary_uncompensated_virtual}. These virtual gate sweep directions are constructed as an arbitrary linear combination of gates $V_1$ to $V_5$. In the second case, we generate compensated virtual gates as per Equation~\ref{eq:compensated_arbitrary_virtual}. In this case, the virtual gate sweep directions are constructed as in the uncompensated case, but a compensation term is included.
Finally, we test the performance of each of the five compensated device gates as per Equation~\ref{eq:single_compensated_gate}. In this case, each virtual gate voltage sweep is that of a single device gate and $V_{P}$. 

For this performance test, a Coulomb peak was first selected by the Bayesian optimisation routine described in Section~\ref{sec:bayesian_optimisation}. The same peak was used for each of the benchmarking runs. To produce the arbitrary virtual gate directions, we generated 100 random vectors from the standard-normal distribution $\vec \epsilon_1, \ldots, \vec \epsilon_{100} \in \mathbb{R}^5$. These vectors were then used to create both uncompensated virtual gates and compensated virtual gates. 

Fig.~\ref{fig:performance} \textbf{a} shows $\sigma_d$ as a function of the length of the gate voltage sweep, $\Delta V_\mathrm{g}$, for the three different sets of gate sweeps described above. 

Without dc compensation, at $\Delta V_g \approx \SI{50}{\milli\volt}$ the Coulomb peak is no longer visible within the measurement window and $\sigma_d$ saturates at \SI{35}{\milli\volt}. Also, as we further increase $\Delta V_g$, a different Coulomb peak becomes visible in the measurement window. 
With dc compensation, we observe that the algorithm drastically reduces $\sigma_d$ over a gate voltage range of \SI{150}{\milli\volt} for both compensated device gates and compensated virtual gates. Ramping compensated device gates has a smaller effect on $\sigma_d$ than ramping compensated virtual gates. This might have to do with non-linearities in the effect of gate voltages that are accentuated in their combination.
Our compensation algorithm applied to $V_3$ allows us to measure device operation regimes from single to double dot for a given Coulomb peak of the sensor dot (Fig.~\ref{fig:performance} \textbf{b}).

\begin{figure*}
\centering

  \includegraphics{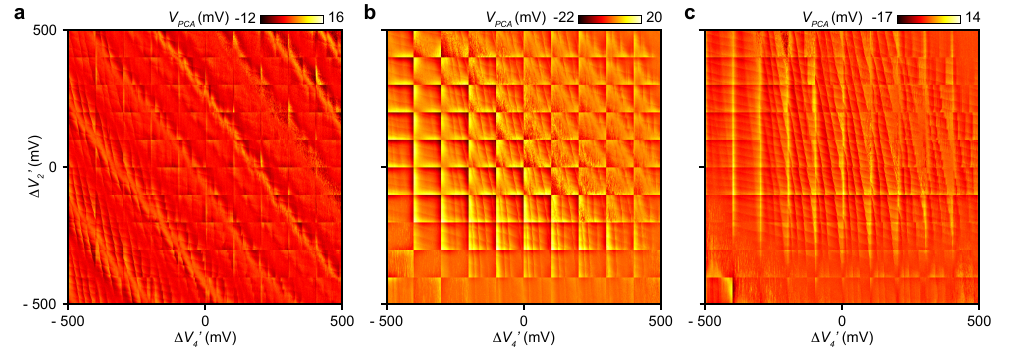}
  \caption{\SI{1000}{\milli\volt} x \SI{1000}{\milli\volt} mosaic of 2D measurement windows for three different cases: \textbf{a} without ac or dc compensation, \textbf{b} just dc compensation for $V_2$ and $V_4$, and \textbf{c} both dc and ac compensation for the same gates. The dc compensation maintains the sensor Coulomb peak offset, \textit{d}, fixed and the ac compensation makes charge transitions significantly clearer. The edges of the 2D measurement windows are still noticeable in panel c due to imperfections in the waveform compensation used to correct the distortion introduced by filters.}\label{fig:large_scan}
\end{figure*}

By combining both ac and dc compensation algorithms, we are able to produce large stability diagrams with clear charge transitions by maintaining the chosen Coulomb peak's offset for the sensor dot fixed. We demonstrate this by measuring \SI{1000}{\milli\volt} x \SI{1000}{\milli\volt} stability diagrams made up of 10 x 10 2D measurement windows. 
In Fig.~\ref{fig:large_scan} (\textbf{a}), it is evident that when no compensation is applied, the visibility of charge transitions in the stability diagram varies significantly. The effect of crosstalk manifests in seven Coulomb peak transitions of the sensor dot.
With dc compensation applied, a Coulomb peak chosen by the Bayesian optimisation routine is used and its offset is kept fixed. The sensor's sensitivity remains thus constant across different 2D measurement windows (Fig.~\ref{fig:large_scan} \textbf{b}). Still, we observe a clear change in the sensor's sensitivity within individual 2D measurement windows. Using both dc and ac compensation, the sensor's sensitivity is kept mostly constant across the full stability diagram (Fig.~\ref{fig:large_scan} \textbf{c}). This shows the extent of our compensation algorithm's capabilities, with charge transitions being visible for almost a volt in each device plunger direction.

\section{Discussion}

We demonstrate how a quantum dot charge sensor can be quickly tuned and compensated for both ac and dc sweeps of a device's gate voltages. A combination of the algorithms presented results in a \SI{1}{\volt} $\times$ \SI{1}{\volt} stability diagram measured while maintaining an optimal transconductance of the sensor dot found by a Bayesian optimiser. 

The gate voltage range within which this compensation approach can be applied is limited by the failure of the linearity assumption underlying the definition of virtual gates. The effectiveness of the compensation therefore reduces as the device gate voltages are swept far away from the gate voltage location at which the virtual gates are calibrated. In our approach, recalibrating the virtual gates takes approximately ten seconds, allowing for the expansion of the gate voltage range in which compensation can be applied. The effect of charge switches and gate voltage drifts that change the sensitivity of the rf-QD can be overcome by integrating a closed-loop feedback system for small measurement windows, such as those demonstrated in Refs.~\cite{Yang2011a, Nakajima2021a}.

The approach we present in this work naturally scales to device architectures with multiple sensors. The ac compensation can also be used for any type of fast 2D measurement, such as those involving interlaced pulse sequences. The algorithm is fully automated and requires no human intervention, and is therefore suitable for use in combination with other automated tuning procedures.

\begin{acknowledgments}
   We thank Nicholas Sim for providing help with the experiment and Sebastian Orbell for helpful discussion. This work was supported by the Royal Society, the EPSRC National Quantum Technology Hub in Networked Quantum Information Technology (EP/M013243/1), Quantum Technology Capital (EP/N014995/1), EPSRC Platform Grant (EP/R029229/1), the European Research Council (Grant agreement 948932), the Scientific Service Units of ISTA through resources provided by the nanofabrication facility and, the FWF-I 05060 and HORIZON-RIA 101069515 projects. 
\end{acknowledgments}

\appendix

\section{Fast two-dimensional measurements}
\label{appendix:measurements}       

We take the fast two-dimensional measurements by applying sawtooth waveforms to gates $V_2$ and $V_4$ using a Tektronix AWG5024 arbitrary waveform generator. These pulses map out the voltage space in a raster pattern with a resolution of $128 \times 128$ pixels and are calibrated to be approximately $100 \times 100$ \si{\milli\volt} dc equivalent after filter compensation. At each pixel, we probe the state of the quantum dot charge sensor using radio-frequency reflectometry. The sensor is illuminated with an rf pulse generated by a Zurich UHFLI at \SI{113.5}{\mega\hertz} at a power of \SI{-90}{\dBm}, close to the resonant frequency of the readout circuit. The   signal is demodulated by the same Zurich UHFLI. The $X$ and $Y$ components of the demodulated signal are sent to an Alazar ATS9440 digitiser where the signal is integrated for \SI{1}{\micro\second}.

\section{Variance of two-dimensional measurements as a score function}
\label{appendix:var}

In this work we use the variance of two-dimensional measurements as a score function for the visibility of the charge sensor in that measurement. In this section we demonstrate why the variance of a two-dimensional measurement with a large number of pixels is insensitive to various sources of experimental noise, and is therefore a reliable metric. We do this by calculating the variance of the variance-based score function. In this section, for the sake of clarity, we refer to the variance-based score function as \textit{the metric}.

Pixels in our measurements are made up of a true signal and a noise function, 

\begin{equation}
M = S + N,
\end{equation}

where $M$ is the measured value,  $S$ is the signal contribution and $N$ is the noise contribution. The noise is drawn from a zero-mean Gaussian distribution with variance $\sigma_N^2$: 

\begin{equation}
N \sim \mathcal{N}(\mu = 0, \sigma_N^2).  
\end{equation}

The variance of our entire two-dimensional measurement, which is the value of the metric, is given by

\begin{equation}
\label{eq:variance_of_measurement}
\sigma_M^2 = \frac{1}{n} \sum_i m_i^2 - \frac{1}{n^2} \left(\sum_i m_i\right)^2, 
\end{equation}

where $n$ is the number of pixels in our measurement $M$ and $i$ indexes each pixel. This formulation makes the assumption that the probability of each value in $M$ is equal.

The variance formula for the propagation of errors is

\begin{equation}
\label{eq:prop_of_vars}
\sigma_f^2 = \left(\frac{\partial f}{\partial x}\right)^2 \sigma_x^2 + \left(\frac{\partial f}{\partial y}\right)^2 \sigma_y^2 + \left(\frac{\partial f}{\partial z}\right)^2 \sigma_z^2 + ... \\
\end{equation}

where $\sigma_f^2$ is the variance of the function $f$ and $\sigma_k^2$ is the variance of variable $k$. Applying Equation~\ref{eq:prop_of_vars} to Equation~\ref{eq:variance_of_measurement} gives us the variance of the metric. We first need the partial derivatives that make up Equation~\ref{eq:prop_of_vars}:

\begin{align}
\frac{\partial \sigma^2_M}{\partial m_j} &= \frac{2 m_i}{n}\delta_{ij} - \frac{2}{n^2} \delta_{ij} \left(\sum_i m_i\right)  \\
&= \frac{2 m_j}{n} - \frac{2}{n^2} \left(\sum_j m_j\right), 
\end{align}

where $j$ indexes the pixels in the measurement $M$. 

Therefore the variance of the metric, $\sigma^2_T$,  is

\begin{align}
\sigma_T^2 &= \sum_j\left(\left(\frac{\partial \sigma_M^2}{\partial m_j} \right) \sigma_{M_j}^2\right)^2, 
\end{align}

The variance of each pixel, $\sigma^2_{M_j}$, is directly from the noise in the measurement and is therefore the same as the variance of $N$, $\sigma_N^2$. Therefore 

\begin{align}
\sigma_T^2 &= \sum_j \sigma_n^2 \left(\frac{2}{n} m_j - \frac{2}{n^2} \left( \sum_j m_j \right) \right)^2 \\
&= \sum_j \frac{4\sigma_n^2}{n^2} \left(m_j - \frac{1}{n} \sum_j m_j \right)^2 \\
&= \frac{4\sigma_n^2}{n^2} \sum_j (m_j - \bar m) \\
&= \frac{4\sigma_n^2}{n} \sum_j \frac{(m_j - \bar m)}{n} \\
&= \frac{4\sigma_n^2}{n} \sigma_M^2 \label{eq:noise_equation}
\end{align}

\begin{figure}
\centering
\includegraphics{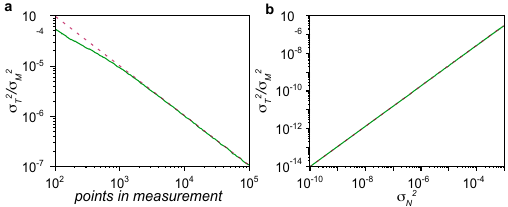}
\caption{Variance of the metric, $\sigma_T^2$ as a function of \textbf{a} number of points in the measurement and \textbf{b} measurement noise scale. Note there were 10,000 points in the measurements for \textbf{b}.}
\label{fig:varvar}
\end{figure}

We can verify this result by simulation. To simulate the charge sensing peak, we created a Lorentzian function centred around the diagonal of our simulated measurement and added noise. By repeatedly sampling the variance of this function we obtained the variance of the variance, which gives the y-axes in Fig.~\ref{fig:varvar}. We then varied the measurement noise scale and the number of points in the measurement. The results agree with those expected due to Equation~\ref{eq:noise_equation}, showing that the variance of this metric is small for a large number of pixels. 

The result is that the noise sensitivity of the variance metric is extremely low, especially when a large number of points are used. Our measurements were $128 \times 128$, or 16,384 pixels. The noise of the metric scales with $\frac{1}{n^2}$, where $n$ is the number of pixels. The noise of this metric scales only linearly with the noise inherent in the measurement. 

\section{Optimising parameters using variance}
\label{appendix:optimising_parameters}

The tuning procedure outlined in the paper typically uses the first principal component, which is the variance, of a two-dimensional measurement to assess whether a regime is good or bad for charge sensing. Large variance means the peak is tall and so highly sensitive. This technique is not restricted to tuning the charge sensor's plunger gate to achieve an optimal position on a Coulomb peak; it is also useful for tuning any experimental parameter to optimise the reflectometry measurement signal-to-noise ratio. 

\begin{figure}[t]
  \includegraphics{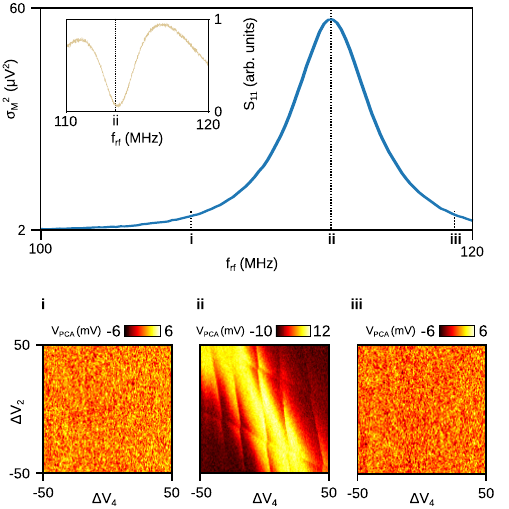}
  \caption{Variance of a \SIrange{100}{100}{\milli\volt} measurement at a fixed point in voltage space as a function of reflectometry frequency. As the frequency approaches the resonant frequency of the readout circuit, the visibility of the charge sensor increases, leading to an increase in variance. The inset shows the cavity $S_{11}$ response as a function of frequency for comparison. Subfigures \textbf{i}, \textbf{ii}, and \textbf{iii} show examples of these measurements at different rf frequencies.}
  \label{fig:frequency_var}
\end{figure}

As an example, consider a measurement that is designed to optimise the frequency of the reflectometry signal. If we sweep the frequency while watching the outcome of successive fast \SIrange{100}{100}{\milli\volt} measurements, we will see that the measurements go from just noise when the frequency does not align with the measurement circuit's resonating frequency into signal when it is aligned and then back to noise when the frequency is too high (Fig.~\ref{fig:frequency_var}). The frequency is optimal when the variance is maximised. 

\section{Principal component analysis}
\label{appendix:pca}

For two data feature dimensions, $x$ and $y$, the data's covariance matrix is defined as 

\begin{equation}
    \Sigma = 
    \begin{bmatrix}
    \sigma(x, x) & \sigma(x, y) \\
    \sigma(y, x) & \sigma(y, y) \\
    \end{bmatrix}
\end{equation}

where $\sigma(x, y)$ represents the covariance of $x$ and $y$.

The paired eigenvectors and eigenvalues of the covariance matrix $\Sigma$ represent the distribution of the dataset's variance in a new orthogonal basis composed of the eigenvectors. Each eigenvalue represents the variance of the data when projected onto its corresponding eigenvector. 

For our measurements, the $x$ and $y$ dimensions are represented by the $I$ (in-phase) and $Q$ (out-of-phase) components of the reflected rf signal. We find the highest eigenvalue of the covariance matrix of these two data sets and take this to be the variance of our data for the score function defined in Section~\ref{sec:score_function}. The paired eigenvector is known as the first principal component of the data. 

The eigenvector is a vector in $IQ$ space with an equation $y = mx + c$. We subtract the mean from the data so the best fit line, which is exactly equivalent to the eigenvector with the highest corresponding eigenvalue, will also have zero mean in both dimensions, meaning its equation simplifies to $y = mx$. The gradient is therefore $m = \frac{y}{x}$, representing an angle of $\arctan{\frac{y}{x}}$. The phase of a data point in XY space is also $\phi = \arctan{\frac{y}{x}}$\cite{VigneauProbing}, so the angle of the best fit line will be at the same angle as that represented by the phase. Projecting the data onto the first principal component is therefore identical to optimally tuning the phase for a given measurement.

\section{Fitting line to Coulomb peak measurement}
\label{appendix:coulomb_fitting}

This section describes how we fit a line to the Coulomb peak for a fast two-dimensional measurements as described in Section~\ref{sec:setup} (e.g.\ Fig.~\ref{fig:d_measurements} \textbf{a}). 

Initially, two linear models are generated from the measurement: the first contains the pixels with the maximum value in each column, and the second with the maximum value in each row. We then generate a line fit for both sets and keep the set with the fit that has the largest $r^2$ value. This model selection routine stabilises the following part of the algorithm. The selected peak line now lies across the top of the Coulomb peak in the charge sensor and has the equation $ax + by + c = 0$ (solid black line,  Fig.~\ref{fig:d_measurements} \textbf{a}). To get a measure of the position of the peak line, we calculate the shortest distance from it to the origin of the measurement $(x_0, y_0)$ (dashed black line, Fig.~\ref{fig:d_measurements} \textbf{a}) and name this distance $d$. The point on the peak line closest to the origin is at $(m, n)$ and the line from here to the origin is perpendicular to the peak line. The distance from $(x_0, y_0)$ to $(m, n)$, is 

\begin{equation}
d = \frac{|ax_0 + by_0 + c|}{\sqrt{a^2 + b^2}}, 
\end{equation}

and since $(x_0, y_0) = (0, 0)$ in pixel space, 

\begin{equation}
d = \frac{|c|}{\sqrt{a^2 + b^2}}.
\end{equation}

\vspace{3mm}

\begin{figure}[ht!]
 \includegraphics{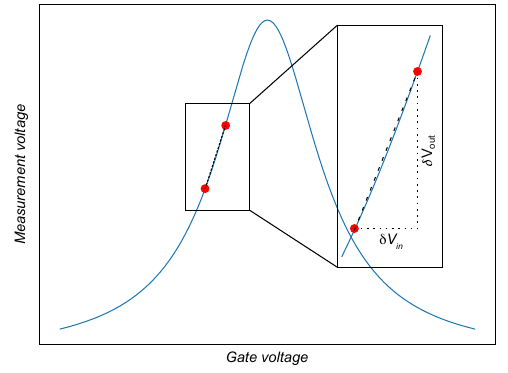}
 \caption{Demonstration of the finite differences method for obtaining the gradient of the sensor peak. $N$ samples are taken at each of the two positions indicated by the red points, and the average of each of these two sets is used to calculate the gradient $\delta V_{\mathrm{out}} / \delta V_{\mathrm{in}}$}
 \label{fig:finite_differences}
\end{figure}

\section{Finite differences gradient method for obtaining coupling strengths}
\label{appendix:finite_differences}

We can obtain the relative coupling strengths of each gate on the potential of the rf-QD by measuring the gradient of a Coulomb peak as a function of each gate's voltage. A Coulomb peak is approximately Lorentzian and therefore does not have a constant gradient. For a local compensation approach, we can approximate the gradient using finite-difference methods.

We initially tune to the side of a Coulomb peak in the rf-QD sensor. For each gate,  we measure at two gate voltages, $V_{\mathrm{gate}} - \frac{\delta V_{\mathrm{in}}}{2}$ and $V_{\mathrm{gate}} + \frac{\delta V_{\mathrm{in}}}{2}$, and calculate the finite difference gradient by $\delta_h [{V_\mathrm{gate}}] = \frac{\delta V_{\mathrm{out}}}{\delta V_{\mathrm{in}}}$. At each of the two gate voltages, we measure $N$ samples and calculate the gradient based on the averages of these samples (Fig.~\ref{fig:finite_differences}).

\bibliographystyle{naturemag}
\bibliography{comp_paper_main}

\begin{thebibliography}{10}
\expandafter\ifx\csname url\endcsname\relax
  \def\url#1{\texttt{#1}}\fi
\expandafter\ifx\csname urlprefix\endcsname\relax\def\urlprefix{URL }\fi
\providecommand{\bibinfo}[2]{#2}
\providecommand{\eprint}[2][]{\url{#2}}

\bibitem{Schoelkopf1998}
\bibinfo{author}{Schoelkopf, R.~J.}, \bibinfo{author}{Wahlgren, P.},
  \bibinfo{author}{Kozhevnikov, A.~A.}, \bibinfo{author}{Delsing, P.} \&
  \bibinfo{author}{Prober, D.~E.}
\newblock \bibinfo{title}{{The radio-frequency single-electron transistor
  (RF-SET): A fast and ultrasensitive electrometer}}.
\newblock \emph{\bibinfo{journal}{Science}} \textbf{\bibinfo{volume}{280}},
  \bibinfo{pages}{1238--1242} (\bibinfo{year}{1998}).

\bibitem{Barthel2010FastDot}
\bibinfo{author}{Barthel, C.} \emph{et~al.}
\newblock \bibinfo{title}{{Fast sensing of double-dot charge arrangement and
  spin state with a radio-frequency sensor quantum dot}}.
\newblock \emph{\bibinfo{journal}{Physical Review B}}
  \textbf{\bibinfo{volume}{81}}, \bibinfo{pages}{161308}
  (\bibinfo{year}{2010}).

\bibitem{Ares2016}
\bibinfo{author}{Ares, N.} \emph{et~al.}
\newblock \bibinfo{title}{{Sensitive Radio-Frequency Measurements of a Quantum
  Dot by Tuning to Perfect Impedance Matching}}.
\newblock \emph{\bibinfo{journal}{Physical Review Applied}}
  \textbf{\bibinfo{volume}{5}}, \bibinfo{pages}{1--6} (\bibinfo{year}{2016}).

\bibitem{Muller2010a}
\bibinfo{author}{M{\"{u}}ller, T.} \emph{et~al.}
\newblock \bibinfo{title}{{An in situ tunable radio-frequency quantum point
  contact}}.
\newblock \emph{\bibinfo{journal}{Applied Physics Letters}}
  \textbf{\bibinfo{volume}{97}}, \bibinfo{pages}{202104}
  (\bibinfo{year}{2010}).

\bibitem{Noiri2020a}
\bibinfo{author}{Noiri, A.} \emph{et~al.}
\newblock \bibinfo{title}{{Radio-Frequency-Detected Fast Charge Sensing in
  Undoped Silicon Quantum Dots}}.
\newblock \emph{\bibinfo{journal}{Nano Letters}} \textbf{\bibinfo{volume}{20}},
  \bibinfo{pages}{947--952} (\bibinfo{year}{2020}).

\bibitem{VigneauProbing}
\bibinfo{author}{Vigneau, F.} \emph{et~al.}
\newblock \bibinfo{title}{{Probing quantum devices with radio-frequency
  reflectometry}}.
\newblock \emph{\bibinfo{journal}{Applied Physics Reviews}}
  \textbf{\bibinfo{volume}{10}}, \bibinfo{pages}{021305}
  (\bibinfo{year}{2023}).

\bibitem{West2019}
\bibinfo{author}{West, A.} \emph{et~al.}
\newblock \bibinfo{title}{{Gate-based single-shot readout of spins in
  silicon}}.
\newblock \emph{\bibinfo{journal}{Nature Nanotechnology}}
  \textbf{\bibinfo{volume}{14}}, \bibinfo{pages}{437--441}
  (\bibinfo{year}{2019}).

\bibitem{Hogg2023}
\bibinfo{author}{Hogg, M.} \emph{et~al.}
\newblock \bibinfo{title}{Single-shot readout of multiple donor electron spins
  with a gate-based sensor}.
\newblock \emph{\bibinfo{journal}{PRX Quantum}} \textbf{\bibinfo{volume}{4}},
  \bibinfo{pages}{010319} (\bibinfo{year}{2023}).

\bibitem{Philips2022}
\bibinfo{author}{Philips, S.~G.} \emph{et~al.}
\newblock \bibinfo{title}{Universal control of a six-qubit quantum processor in
  silicon}.
\newblock \emph{\bibinfo{journal}{Nature}} \textbf{\bibinfo{volume}{609}},
  \bibinfo{pages}{919--924} (\bibinfo{year}{2022}).

\bibitem{Fedele2021}
\bibinfo{author}{Fedele, F.} \emph{et~al.}
\newblock \bibinfo{title}{Simultaneous operations in a two-dimensional array of
  singlet-triplet qubits}.
\newblock \emph{\bibinfo{journal}{PRX Quantum}} \textbf{\bibinfo{volume}{2}},
  \bibinfo{pages}{040306} (\bibinfo{year}{2021}).

\bibitem{Yang2011a}
\bibinfo{author}{Yang, C.~H.}, \bibinfo{author}{Lim, W.~H.},
  \bibinfo{author}{Zwanenburg, F.~A.} \& \bibinfo{author}{Dzurak, A.~S.}
\newblock \bibinfo{title}{{Dynamically controlled charge sensing of a
  few-electron silicon quantum dot}}.
\newblock \emph{\bibinfo{journal}{AIP Advances}} \textbf{\bibinfo{volume}{1}}
  (\bibinfo{year}{2011}).

\bibitem{Nakajima2021a}
\bibinfo{author}{Nakajima, T.} \emph{et~al.}
\newblock \bibinfo{title}{{Real-Time Feedback Control of Charge Sensing for
  Quantum Dot Qubits}}.
\newblock \emph{\bibinfo{journal}{Physical Review Applied}}
  \textbf{\bibinfo{volume}{15}}, \bibinfo{pages}{1} (\bibinfo{year}{2021}).

\bibitem{Hensgens2017}
\bibinfo{author}{Hensgens, T.} \emph{et~al.}
\newblock \bibinfo{title}{{Quantum simulation of a Fermi-Hubbard model using a
  semiconductor quantum dot array}}.
\newblock \emph{\bibinfo{journal}{Nature}} \textbf{\bibinfo{volume}{548}},
  \bibinfo{pages}{70--73} (\bibinfo{year}{2017}).

\bibitem{Botzem2018}
\bibinfo{author}{Botzem, T.} \emph{et~al.}
\newblock \bibinfo{title}{{Tuning Methods for Semiconductor Spin Qubits}}.
\newblock \emph{\bibinfo{journal}{Physical Review Applied}}
  \textbf{\bibinfo{volume}{10}}, \bibinfo{pages}{1} (\bibinfo{year}{2018}).

\bibitem{Volk2019}
\bibinfo{author}{Volk, C.} \emph{et~al.}
\newblock \bibinfo{title}{{Loading a quantum-dot based “Qubyte” register}}.
\newblock \emph{\bibinfo{journal}{npj Quantum Information}}
  \textbf{\bibinfo{volume}{5}}, \bibinfo{pages}{1--12} (\bibinfo{year}{2019}).

\bibitem{Mills2019}
\bibinfo{author}{Mills, A.~R.} \emph{et~al.}
\newblock \bibinfo{title}{{Shuttling a single charge across a one-dimensional
  array of silicon quantum dots}}.
\newblock \emph{\bibinfo{journal}{Nature Communications}}
  \textbf{\bibinfo{volume}{10}} (\bibinfo{year}{2019}).

\bibitem{Jirovec2021}
\bibinfo{author}{Jirovec, D.} \emph{et~al.}
\newblock \bibinfo{title}{{A singlet-triplet hole spin qubit in planar Ge}}.
\newblock \emph{\bibinfo{journal}{Nature Materials}}
  \textbf{\bibinfo{volume}{20}}, \bibinfo{pages}{1106--1112}
  (\bibinfo{year}{2021}).

\bibitem{Stehlik2015}
\bibinfo{author}{Stehlik, J.} \emph{et~al.}
\newblock \bibinfo{title}{{Fast Charge Sensing of a Cavity-Coupled Double
  Quantum Dot Using a Josephson Parametric Amplifier}}.
\newblock \emph{\bibinfo{journal}{Phys. Rev. Applied}}
  \textbf{\bibinfo{volume}{4}}, \bibinfo{pages}{14018} (\bibinfo{year}{2015}).

\bibitem{Schupp2020}
\bibinfo{author}{Schupp, F.~J.} \emph{et~al.}
\newblock \bibinfo{title}{{Sensitive radiofrequency readout of quantum dots
  using an ultra-low-noise SQUID amplifier}}.
\newblock \emph{\bibinfo{journal}{Journal of Applied Physics}}
  \textbf{\bibinfo{volume}{127}} (\bibinfo{year}{2020}).

\bibitem{vanStraaten2022a}
\bibinfo{author}{van Straaten, B.} \emph{et~al.}
\newblock \bibinfo{title}{All rf-based tuning algorithm for quantum devices
  using machine learning}.
\newblock \eprint{arXiv:2211.04504}.

\end{thebibliography}

\end{document}